\documentclass[%
preprint,
nofootinbib, nobibnotes,
amsmath,amssymb, aps, prd,
]{revtex4-2}
\usepackage{hyperref}

\usepackage{amsmath}
\usepackage{amssymb}
\usepackage{amsthm}
\usepackage{bm}

\usepackage{graphicx,color}

\def\be{\begin{equation}}
\def\ee{\end{equation}}
\def\bea{\begin{eqnarray}}
\def\eea{\end{eqnarray}}
\def\ptl{\partial}

\begin{document}

\centerline{\large{ \bf Illusiveness of the problem of time}}
\bigskip
\bigskip
\centerline{S.L. Cherkas $^{a,}\footnote{Corresponding author}$,
V.L. Kalashnikov$^b$}
\bigskip
\bigskip
\centerline{$^a$Institute for Nuclear Problems, Belarus State
University} \centerline{Minsk 220030, Belarus} \centerline{Email:
cherkas@inp.bsu.by}
\medskip
\centerline{$^b$Dipartimento di Ingegneria dellInformazione, Elettronica e Telecomunicazioni,} \centerline{ Sapienza Universit\'a di Roma, 00189 - Roma, RM, Italia} \centerline{Email:
vladimir.kalashnikov@uniroma1.it}

\bigskip
\centerline{Submitted: March 30, 2020}
\bigskip
{
The essay is devoted to the problem of time in the context of quantum cosmology, which acquires a philosophical level to date. At an example of the minisuperspace model, we show that this problem is illusive in the sense that it does not prevent to calculate mean values of the operators over the quantum state of the universe. Contrariwise, the different approaches to the description of these time-dependent mean values give similar results. 
}
\bigskip
\bigskip

\centerline {\it Essay written for the Gravity Research Foundation
2020.}

\newpage

 \noindent
\textit{\textbf{View on 'time'}}

Although the different methods for the calculations of operator mean values in quantum cosmology \cite{arxiv2020} will be discussed below, a few general remarks about time are required. According to Proclus, time is a pure duration, an absolute and continuous fluidity. If one would like to describe the mobility of something, she immediately notices that it always depends on some other thing, and its moving should be explained by a third thing and so on. Consequently, to explain the motion of all things without the need of an infinite chain of 'movers', one needs to accept that there is such a thing that moves itself and moves all the other things. It is an eternal entity of 'time': 'Time  is eternal not merely in its essence, but it is also ever same in its inner activity, and it is only so far as it is participated by things outside itself that is movable' \cite{Proclus}. As it was properly noticed by Martin Heidegger \cite{Heidegger}\footnote{Time has been discussed  in many philosophical books, but we mention only two of them.}, one could carefully dismantle different clocks and chronometers, but he will not find "time" in them.

Consequently, 'time' lies under the world and physics, but still, it should be an instrumental concept, allowing to describe phenomena in a very economical way, probably, more economical than 'relational dynamics.' It would be rather strange to talk about the 'emergence' of 'time,' since it is an under-everything entity. It is more convenient to consider that it always exists and permeates all the classical and quantum phenomena. The ability to choose time variables differently shows simply that in mathematical terms, 'time' is an equivalence class. On the other hand, if one wants to shrink the time into one point, she comes to 'eternity' \cite{Proclus}, which is the second side of the coin.

Here we will come from the assumption that some particularly chosen gauge means some ideal particular clock. Another hypothesis is that the gravity in the form of general relativity is a kind of conventional systems with constraints, and one could use ordinary rules of quantization of these systems \cite{ruffini2005}.

\textit{\textbf{Classical picture}}

Let us consider Lagrangian of minisuperspace model for gravity and a real massless scalar
field  $\phi$:
\be
L=\frac{1}{2N}\left(-{M_p^2}a^{\prime
2}+a^2\phi^{\prime 2}\right),
\label{lag1}
\ee
which corresponds to uniform,
isotropic and  flat
universe
\be
ds^2=g_{\mu\nu}dx^\mu dx^\nu=a^2(N^2d\eta^2-d^2\bm r).
\ee
The reduced Planck mass $M_p=\sqrt{\frac{3}{4\pi G}}$ will be set to unuity
further. 
Corresponding 
Hamiltonian
\be
H=N\left(-\frac{1}{2}p_a^2+\frac{\pi_\phi^2}{2a^2}\right),
\ee
 is also Hamiltonian constraint $\Phi_1=H=0$
due to $\frac{\ptl  L}{\ptl N}=0$.

Solution of the equations of motion is
\be
a=\sqrt{2|\pi_\phi| \eta},
~~~~~~~~~~~~\phi=\frac{\pi_\phi}{2|\pi_\phi|}\ln\eta +const.
\label{at}
\ee

According to Eq. (\ref{at}) a gauge fixing condition
\be
\Phi_2=a-\sqrt{2|\pi_\phi| \eta}=0
\label{f2}
\ee
which conserved in time 
could be introduced
in addition to the constraint $\Phi_1$. 

One can see explicit time evolution under some particular gauge
fixing. Moreover, for this simple example, the system could be reduced
to the only degree of freedom \cite{Barv2014,Kamen2019}.

Let us take $\pi_\phi$ and $\phi$ as the physical variables, then
$a$ and $p_a$ have to be excluded by the constraints. Substituting
$p_a$, $a^\prime$ and $a$ into (\ref{lag1}) one comes to
\be
L=\int \left(\pi_\phi\phi^\prime
-H_{phys}(\phi,\pi_\phi,\eta)\right)d\eta,
\label{L2}
\ee
where
\be
H_{phys}(\phi,\pi_\phi,\eta)=p_a
a^\prime=\frac{|\pi_\phi|}{2\,\eta}.
\label{hph}
\ee

\textit{\textbf{Quantum pictures}}

The most simple and straightforward way to the description of the quantum evolution is based on the Schr\"{o}odinger equation \cite{Barv2014,Kamen2019,arxiv2020}
\be
i\ptl_\eta \Psi=\hat H_{phys}\Psi
\label{sh}
\ee
with the physical Hamiltonian (\ref{hph}). In the momentum
representation, the operators become
\be
\hat \pi_\phi=k, ~~~~~\hat \phi=i\frac{\ptl}{\ptl k}.
\label{mom}
\ee

Solution of Eq. (\ref{sh}) is written as
\be
\Psi(k,\eta)=C(k)|2 k\eta|^{-i|k|/2}e^{i|k|/2},
\label{wave}
\ee
where $C(k)$ is momentum wave packet.  An arbitrary operator $\hat A$
 build from $\hat
\phi=i\frac{\ptl}{\ptl k}$ and $a=\sqrt{2|k|\eta}$ is, in fact, the function of $\eta$, $k$, $i\frac{\ptl}{\ptl k}$.
Using the wave  function (\ref{wave}) it is possible to calculate its mean value ({\it method A})
\be
<C|\hat A|C>=\int \Psi^*(k,\eta)\hat A\,\Psi(k,\eta)dk.
\ee

Because the base wave functions $\psi_k=|2 k\eta|^{-i|k|/2}e^{i|k|/2}$ contain the module of $k$, a singularity may arise at $k=0$ if $\hat A$
 contains degrees of differential operator $\frac{\ptl}{\ptl k}$. That may violate hermicity. To avoid this, the wave packet $C(k)$ has turn to zero at $k=0$. For instance, it could be taken in the Gaussian function multiplied by $k^2$.
\be
C(k)=\frac{4 \sigma^{5}}{3\sqrt \pi}k^2\exp\left(-\frac{k^2}{2\sigma^2}\right).
\label{pak}
\ee
 Let us come
to calculation of the concrete mean values of the operators $\hat \phi^2$
and $a$ taking $\sigma=1$ for simplicity:
\bea
<C|a|C>=\frac{4}{3} \sqrt{\frac{2}{\pi
}}\sqrt{\eta}\int_{-\infty}^\infty e^{-k^2} k^{9/2} dk=\frac{4}{3}
\sqrt{\frac{2}{\pi
}}\Gamma(11/4)\sqrt{\eta},~~~~~~~~~~~~~~~~~~~\label{resa}\\
<C|\hat\phi^2|C>=\frac{1}{3\sqrt{\pi}}\int_{-\infty}^\infty
e^{-k^2}\biggl(-4 k^6+k^4 \left(20+\ln ^22+\ln (\eta  \left|
k\right| ) \ln (4 \eta \left| k\right| )\right)\nonumber-\\8 k^2+2
i \left| k\right| ^3 \left(-2 k^2 \ln (2 \eta \left| k\right| )+4
\ln (\eta  \left| k\right| )+4\ln
   2+1\right)\biggr)dk=\nonumber\\
   \frac{1}{12} \ln \eta  (3 \ln \eta -3 \gamma +8)+\frac{\pi ^2}{32}+\frac{\gamma ^2}{16}-\frac{\gamma
   }{3}+\frac{4}{3},~~~~~~~~~~~~
   \label{resphi2}
\eea
 where $\Gamma$ is the Gamma function, and $\gamma$ is the Euler constant. It could be noted that the imaginary part in (\ref{resphi2}) disappears after integration on $k$ due to
 hermicity of $\hat \phi$.

The problem of time began from the discussing of the Wheeler-DeWitt
equation (WDW) equation \cite{Wheel67,DeWitt67,Barv2014,Shest2018} which could be considered as a ``mathematical implementation of eternity''. It is often stated that the WDW equation
\be
\left(\frac{\ptl^2}{\ptl\alpha^2}-\frac{\ptl^2}{\ptl\phi^2}\right)\Psi(\alpha,\phi)=0,
\label{WDW1}
\ee
where $\alpha=\ln a$,
does not contain time explicitly. Indeed, it is true. Then, it is usually stated that the WDW equation forbids time evolution. Certainly, it is wrong, if one considers full quantum picture,
including gauge fixing and evaluation of the mean values of the operators. The point is that the WDW equation has to be supplemented by a scalar product.

Scalar products for the Klein-Gordon equation are discussed in \cite{mostf}, where "current" and "density" products were suggested. Here we will use an only scalar product of the "current" type including the hyperplane $\alpha=\alpha_0$ which results in the following formula
for a mean value of some operator \cite{mostf,arxiv2020}
\be
  <\Psi|\hat A|\Psi>=i \int \biggl( \Psi^* {\hat D^{1/4}}\hat A\,{\hat
  D^{-1/4}}\frac{\ptl \Psi}{\ptl
\alpha }-\left(\frac{\ptl
\Psi^*}{\ptl \alpha }\right){\hat D^{-1/4}}\hat A\,{\hat
D^{1/4}}\,\Psi\biggr)\biggr|_{\,\alpha=\alpha_0} d\phi
,\label{mean3}
\ee
where the operator $\hat D=-\frac{\ptl^2}{\ptl\phi^2}$. In the
momentum representation (\ref{mom}) the WDW equation (\ref{WDW1}) looks as
\be
\left(\frac{\ptl^2}{\ptl\alpha^2}+k^2\right)\psi(\alpha,k)=0,
\ee
and due to $\hat D^{1/2}=|k|$ the scalar product (\ref{mean3}) takes the
form.
\be
<\Psi|\hat A|\Psi>=i \int C^*(k)e^{i|k|\alpha}\hat A
e^{-i|k|\alpha}C(k)\biggr|_{\,\alpha=\alpha_0}dk,
\label{meanf}
\ee
where
\be \Psi(\alpha,\phi)=\int e^{ik\phi}
\psi(\alpha,k)dk=\int
\frac{e^{ik\phi-i|k|\alpha}}{\sqrt{2|k|}}C(k)dk
\ee is taken. To introduce the time evolution into this picture one
has to take a time-dependent integration plane in (\ref{meanf})
instead of $\alpha=\alpha_0$ by writing
$\alpha=\frac{1}{2}\ln\left(2|k|\eta\right)$ according to
(\ref{f2}), i.e., to $\Phi_2=0$.

However, if the
operator $\hat A(\alpha,k,i\frac{\ptl}{\ptl
\alpha},i\frac{\ptl}{\ptl k})$ contains differentiations
$\frac{\ptl}{\ptl k}$ or $\frac{\ptl}{\ptl \alpha}$, hermicity
could be lost. To prevent this, let us rewrite (\ref{mean3}),
(\ref{meanf}) in the form ({\it method B}) \cite{arxiv2020}
\bea
<\psi|\hat A|\psi>= \int \psi^*(\alpha,k)\biggl(|k|^{1/2}\hat
A|k|^{-1/2}\delta(\alpha -\frac{1}{2}\ln(2|k|\eta))\hat
p_\alpha+~~~~~~~~\nonumber\\\hat p_\alpha\delta(\alpha
-\frac{1}{2}\ln(2|k|\eta))|k|^{-1/2}\hat
A|k|^{1/2}\biggr)\psi(\alpha,k) d \alpha dk,
\label{sc2}
\eea
where $p_\alpha=i\frac{\ptl}{\ptl \alpha}$ and  hermicity of $\hat A$ relative $\alpha,k$ variables are
implied. In this case, no problem with hermicity arises if one takes the functions $\psi(\alpha,k)$ turning to zero at $\alpha\rightarrow \pm \infty$ to provide throwing over differential operators $\ptl/\ptl \alpha$ over integration by parts. The functions $\psi(\alpha,k)=\frac{e^{-i|k|\alpha}}{\sqrt{2|k|}}C(k)$ do not poses such a property, thus, we shall take the functions
\be
\psi(\alpha,k)=\frac{e^{-i|k|\alpha-\alpha^2/\Delta}}{\sqrt{2|k|}}C(k)
\label{ftest}
\ee
in the intermediate calculations and, then, after integration over
$\alpha$, tends $\Delta$ to infinity. 
Another version with the anticommutative variables could be
suggested in the form of ({\it method C}) \cite{arxiv2020}
\bea
<\psi|A|\psi>= \int
\psi^*(\alpha,k)\exp\biggl(i\lambda\bigl(\alpha
-\frac{1}{2}\ln(2|k|\eta)\bigr)+\bar \theta\theta
\hat p_\alpha+~~~~~~~~~~~\nonumber\\\frac{1}{2}\bar \chi\chi\left(|k|^{-i/2} \hat A\,|k|^{i/2}+|k|^{i/2} \hat A\,|k|^{-i/2}\right)\biggr)\psi(\alpha,k)d\lambda
d\alpha  dk d \theta d\bar\theta d \chi d\bar \chi,
\label{sc22}
\eea
where anticommutating Grassman variables $\theta_i=(\theta,\chi)$, $\bar \theta_i=(\bar\theta,\bar\chi)$ have  been
introduced. Again, for reasons of hermicity, we take the functions 
(\ref{ftest}) for calculations.

Another approach to describe the time evolution is to take the classical
equations of motion and then quantize them ({\it method D}), i.e., write ``hat'' under every quantity \cite{Cher2005,Cher2012,Cher2013,Cher2017}. Then, the operator equations of motion take the form \cite{arxiv2020}:
\be
\hat \phi^{\prime\prime}+2\hat \alpha^\prime\hat
\phi^\prime=0,~~~~~~~ \hat \alpha^{\prime\prime}+\hat
\alpha^{\prime 2}+\hat \phi^{\prime 2}=0.
\label{eqop}
\ee
One needs to find the commutation relations of the operators $\hat
\alpha(\eta)$, $\hat \phi(\eta)$. The problem was solved by Dirac, who has introduced the Dirac brackets for the system with constraints
postulating that the commutator relations of the operators have to be analogous to their Dirac brackets. However, it is not always
possible to find operator realization of this commutator
relations. The quasi-Heisenberg picture \cite{Cher2005,Cher2012,Cher2013,Cher2017} assumes finding the operator
realization only at the initial moment of time and then allow
operators to evolve according to the equations of motion. The
initial conditions for operators could be taken in the form
\be
\hat \alpha(0)=\alpha_0, ~~~ \hat
\alpha^\prime(0)=e^{-2\alpha_0}|k|,~~~\hat
\phi(0)=i\frac{\ptl}{\ptl k}, ~~~\hat
\phi^\prime(0)=e^{-2\alpha_0}k.
\label{in}
\ee

The solution of the operator equations of motion (\ref{eqop}) with the
initial conditions (\ref{in}) are
\be
\alpha(\eta)=\alpha_0+\frac{1}{2}\ln\left(1+2|k|\eta\,
e^{-2\alpha_0}\right),~~~\hat \phi(\eta)=i\frac{\ptl}{\ptl
k}+\frac{k}{2|k|}\ln\left(1+2|k|\eta \,e^{-2\alpha_0}\right).
\ee

To built the Hilbert space, where these quasi-Heisenberg operators
act, one may use the WDW equation (\ref{WDW1}) and the scalar product
(\ref{meanf}), but the value of $\alpha_0$ should be set to minus infinity \cite{Cher2005,Cher2012,Cher2013,Cher2017} at the end of an evaluation, i.e., $\alpha_0\rightarrow -\infty$, which corresponds to $a\rightarrow0$ at $\eta=0$. The explicit calculation gives the same mean values as (\ref{resa}), (\ref{resphi2}).

One more exciting picture suggests unconstrained dynamics ({\it method E}). It is believed \cite{Kaku2012,Savchenko2004,Vereshkov2013}, that the Grassman variables allow obtaining the Hamiltonian 
\be
H=N\left(-\frac{1}{2}p_a^2+\frac{\pi_{\phi}^2}{2a^2}\right)+\frac{1}{N}\pi_\theta\pi_{\bar \theta},
\label{ham2}
\ee
which drives unconstrained dynamics. Results of the mean values calculation is presented in Table \ref{tab}. The mean value of $<C|a^2|C>$ turns out to be the same for all the methods considered \cite{arxiv2020}.  For the {\it method E}, we were not able to calculate the mean values of the other operators for two reasons: because we use the most primitive way of calculation by expanding the exponent $e^{-i\hat H\eta}$  in degrees of $\eta$ and utilize the primitive regularisation
under transition from the extended space (see e.g. \cite{Mont}) to the space of the WDW equation solutions. 

\begin{figure}[ht]
  \includegraphics[width=8cm]{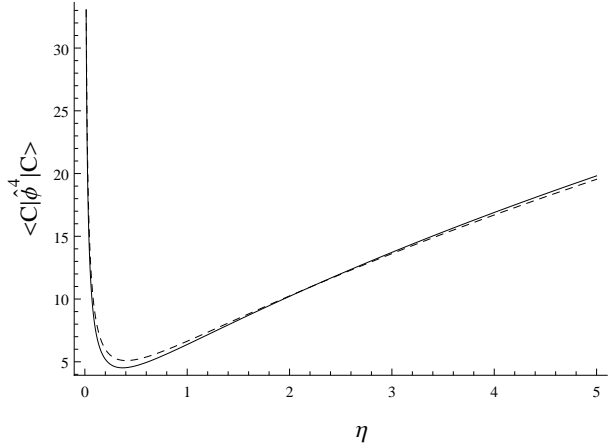}\\
  \caption{Mean value of $\hat \phi^4$ over wave packet (\ref{pak}) for the methods A, D- solid line, and methods B, C- dashed line. 
   }\label{fig2}.
\end{figure}

\begin{table}
\caption{Comparison of the mean values calculated by the different methods. Capital letters denote a method. Plus implies that values obtained by the different methods coincide. Crosses of two types in a circle mean that the values obtained at least two different methods coincide.
}
\label{tab}
\vspace{7pt}
\begin{tabular}{llllll}
\hline
Method & A & B & C & D & E\\
\hline
$a$ & + & + & + & + &  \\
$a^2$ &+&+ &+ & +&+\\
$\hat \phi^2$ &+ &+ &+&+ & \\
$\hat \phi^4$ &$\oplus$ &$\otimes$ &$\otimes$& $\oplus$&\\
$\hat \phi^6$ &$\oplus$ & & &$\oplus$ & \\
$a\hat \phi^2 +\hat\phi^2 a$~~~~~ &$\oplus$ & & &$\oplus$& \\\hline
\end{tabular}
\vspace*{-2pt}
\end{table}

The  methods {\it A,B,C,D} produce the same value of the operators $a$, $\phi^2$ as it is shown in Table \ref{tab}. For the mean value of $\hat \phi^4$, some difference emerges as it is shown in Fig. \ref{fig2}. It is not the uncertainty of numerical calculations because they are fully analytical and have been performed using Mathematica \cite{arxiv2020}. However, let us emphasize that it does not mean that the different methods are nonequivalent. Generally, different methods should not have the same Hilbert
space when producing the same values of the different operators. Only correspondence between these spaces should exist, i.e., these spaces have be connected by some transformation.

\textit{\textbf{Conclusion}}

As one could see that the description of quantum evolution is very
straightforward and unambiguous but teems with different details
such as choosing a scalar product and an operator ordering which are
typical for quantization of the systems with constraints \cite{ruffini2005}.

It is shown that if one wants to discuss the quantum evolution of the
universe by calculating the mean values of the operators, she has no serious obstacles for this. Namely, the
"problem of time" \cite{Kuchar1991,Isham1993,Shest2004,rovelli2009,anderson2010} does not exist as a real problem.

The WDW equation tells nothing about the time evolution without determining the scalar product so that this equation alone is only halfway to a full quantum picture. 

\textit{\textbf{Acknowledgments}}

S.L.C. is grateful to Dr. Tatiana Shestakova for discussions.

\bibliography{time}

\end{document}